\documentclass[preprint,preprintnumbers,aps,prd,floatfix,superscriptaddress,nofootinbib]{revtex4-1}
\usepackage{booktabs}
\usepackage{dcolumn}
\usepackage{bm}
\usepackage[usenames ,dvipsnames]{xcolor}
\usepackage{slashed}
\usepackage{graphicx}
\usepackage{placeins}
\usepackage[bookmarksnumbered, pdfstartview=FitH,colorlinks,urlcolor=blue, citecolor=blue,linkcolor=blue] {hyperref}
\usepackage{appendix}
\usepackage[utf8]{inputenc}
\usepackage{amsmath}
\begin{document}
\title{\texorpdfstring{A Class of Exact Single-Field Inflationary\\ Solutions beyond Slow Roll}{A Class of Exact Single-Field Inflationary Solutions beyond Slow Roll}}
\author{Jia-Wei Zhang}
\email{Corresponding author: jwzhang@cqust.edu.cn}
\affiliation{Department of Physics, Chongqing University of Science and Technology, Chongqing, 401331,
 China}
\author{Bai-Cian Ke}
\email{Corresponding author: baiciank@ihep.ac.cn}
\affiliation{School of Physics, Zhengzhou University, Zhengzhou, Henan 450001, China}
\author{Yao Yu}
\email{Corresponding author: yuyao@cqupt.edu.cn}
\affiliation{Chongqing University of Posts \& Telecommunications, Chongqing, 400065, China}
\affiliation{Department of Physics and Chongqing Key Laboratory for Strongly Coupled Physics, Chongqing University, Chongqing 401331, People's Republic of China}
\author{Dong-Ze He}
\affiliation{Chongqing University of Posts \& Telecommunications, Chongqing, 400065, China}
\author{Shou-Jia Wang}
\affiliation{Chongqing University of Posts \& Telecommunications, Chongqing, 400065, China}

\begin{abstract}
We construct exact solutions for single-field inflaton dynamics without invoking the slow-roll approximation. A suitable change of variables reduces the background equation to an Abel equation of the first kind. Although a generic Abel equation is not analytically solvable, we identify a class of inflaton potentials for which the transformed equation admits exact solutions. The resulting framework contains constant-roll inflation as a special case and also accommodates solutions with a constant second Hubble-flow parameter. We analyze the linear local attractor behavior and superhorizon evolution of these rolling backgrounds. Using the public joint CMB likelihood contours in the $(n_s,r)$ plane, we identify compatible parameter regions and show that one rolling branch can also yield $50\leq N_*<60$. Direct numerical evolution of the scalar and tensor modes at representative points validates the local-index predictions to better than $7\times10^{-4}$ in $n_s$ and $2\times10^{-6}$ in $r$. The exact family extends beyond slow roll, although the observationally selected regions displayed here lie close to the slow-roll regime.
\end{abstract}
\maketitle
\section{Introduction}
Cosmic inflation \cite{Guth:1980zm} provides a compelling framework for the early Universe and addresses fundamental shortcomings of the standard hot big-bang scenario, including the horizon and flatness problems \cite{Starobinsky:1980te,Linde:1983gd}. In the simplest realization, the potential energy of a scalar field, the inflaton, dominates the total energy density and drives a period of accelerated expansion.

The background dynamics are governed by nonlinear differential equations and are therefore commonly treated using analytical approximations. The slow-roll approximation is the standard example: the inflaton evolves sufficiently slowly along a relatively flat potential that its acceleration can be neglected. Slow roll is sufficient for inflation, but it is not necessary. Exact solutions outside this regime are consequently valuable both as controlled examples of non-slow-roll dynamics and as benchmarks for approximate or numerical treatments.

In this work, we develop a systematic method for obtaining exact solutions of the inflaton equation. A suitable change of variables maps the background dynamics to an Abel equation of the first kind, which becomes exactly solvable for a specific class of potentials. The construction encompasses constant-roll inflation \cite{Motohashi:2014ppa}, a well-studied departure from the conventional slow-roll scenario \cite{Yi:2017mxs,Motohashi:2017aob,Motohashi:2019rhu,Anguelova:2017djf,Karam:2017rpw,Ito:2017bnn,Tasinato:2023ukp,Tasinato:2020vdk,Cicciarella:2017nls,Gao:2018tdb,Cai:2016ngx,Inui:2024sce,Liu:2024uiy,Motohashi:2025qgd}, while extending it to a broader family of exact non-slow-roll backgrounds. We then examine the linear local attractor behavior, superhorizon evolution, and observational predictions of these rolling solutions.

Relative to earlier Abel-equation formulations of scalar-field
cosmology \cite{Muslimov:1990be,Yu:2025flh}, the new ingredients are
the solvable rational ansatz for $f(y)$ in Eq.~(\ref{tuy1}), the
resulting two-parameter family of exact potentials, and its
classification into dynamically distinct rolling branches.  We also
show that the same local coefficient governs convergence of the field
velocity and decay of the time-dependent superhorizon mode, and we
combine these local conditions with explicit $(n_s,r)$ and e-fold
scans, and validate representative spectra by direct mode evolution.
These steps, rather than the Abel transformation by itself,
constitute the main extension developed here.

The remainder of the paper is organized as follows. In Sec.~\ref{sec:kinematics} we construct the exact solutions. Section~\ref{sec:physical significance} analyzes their physical properties, including the local attractor behavior of the rolling backgrounds and the evolution of cosmological perturbations. We then present the numerical constraints and summarize our conclusions.

\section{Construction of Exact Solutions}\label{sec:kinematics}

In a spatially flat Friedmann--Lema\^{\i}tre--Robertson--Walker spacetime, a canonical homogeneous inflaton field $\phi$ obeys the Klein--Gordon equation
\begin{eqnarray}\label{kg}
\ddot{\phi}+3H\dot{\phi}+\frac{dV}{d\phi} &=& 0,
\end{eqnarray}
where overdots denote derivatives with respect to cosmic time $t$, $V(\phi)$ is the inflaton potential, and $H\equiv\dot{a}/a$ is the Hubble parameter, with $a$ being the scale factor. When the Universe is dominated by the inflaton field, the Friedmann equation takes the form
\begin{eqnarray}\label{friedmann}
H^2 &=& \frac{1}{3M_p^2}\left(\frac{1}{2}\dot{\phi}^2+V(\phi)\right),
\end{eqnarray}
where $M_p\equiv1/\sqrt{8\pi G}$ is the reduced Planck mass.

The background dynamics of a canonical single-field inflationary model can then be recast as an Abel equation of the first kind,
\begin{eqnarray}\label{zd}
  \frac{dy}{d\phi} &=& f_3 y^3 + f_2 y^2 + f_1 y + f_0 = (y^2 - 1)\left(\frac{y}{2} \frac{d\ln V}{d\phi} + \frac{\sqrt{6}}{2M_p}\right)
\end{eqnarray}
where \( f_3 = -f_1 = \frac{1}{2} \frac{d \ln V(\phi)}{d\phi} \), \( f_2 = -f_0 = \frac{\sqrt{6}}{2M_p} \), and \( y = \frac{\sqrt{\dot{\phi}^2 + 2V(\phi)}}{\dot{\phi}} \). Here \(M_p\) denotes the reduced Planck mass. Equation~(\ref{zd}) is an Abel equation of the first kind \cite{er2}; a closely related formulation of scalar-field dynamics was given in Ref.~\cite{Muslimov:1990be,Yu:2025flh}.

The corresponding background quantities can be written as
\begin{eqnarray}\label{zy1}
  \dot{\phi} &=& \frac{y}{|y|} \sqrt{\frac{2V}{y^2 - 1}}, \quad H = \frac{|y|}{\sqrt{3}M_p} \sqrt{\frac{V}{y^2 - 1}}
\end{eqnarray}

As a benchmark, the familiar power-law inflationary solution \cite{Yu:2025flh,er7} follows from a special form of the Abel equation; the relevant solution is discussed in Eq.~(1.51) of Ref.~\cite{er2} and in Ref.~\cite{er4}.

More generally, Eq.~\ref{zd} provides a constructive route to particular analytical solutions. We regard \( f(y) = \frac{d \ln V}{d\phi} \) as a prescribed function of \( y \). Substitution into Eq.~\ref{zd} yields a separable equation from which the relation between \( y \) and \( \phi \) can be determined, preferably in explicit form. Inserting that relation back into \( f(y) = \frac{d \ln V}{d\phi} \) then determines the corresponding potential.

The central step is therefore the choice of \( f(y) \). As one particular choice among many other possibilities, we take
\begin{eqnarray}
  \frac{d \ln V}{d \phi} = f(y) &=& \frac{1}{M_p}\left[\frac{\bar{\alpha}}{y} + \frac{\lambda}{y - 1} + \frac{\lambda}{y + 1}\right]
\end{eqnarray}
where \( \bar{\alpha}, \lambda \) are arbitrary constants. Substituting this ansatz into Eq.~\ref{zd} gives
\begin{eqnarray}\label{tuy1}
  \frac{dy}{d\phi} &=& \frac{1}{2M_p}\left[ (\alpha + 2\lambda)y^2  - \alpha \right]
\end{eqnarray}
where \( \alpha = \bar{\alpha} + \sqrt{6} \). For later convenience, we rewrite Eq.~\ref{tuy1} as follows. The degenerate case \( f(y) \equiv 0 \), obtained for \( \bar{\alpha} = \lambda = 0 \), is excluded because the relations below no longer apply.
\begin{eqnarray}\label{hx1}
  \frac{d\varpi}{d\phi} &=& \frac{1}{M_p} \left(\frac{\alpha}{2\sqrt{3}}\varpi^2  - \frac{\sqrt{3}(\alpha + 2\lambda)}{2}\right)
\end{eqnarray}
\begin{eqnarray}\label{hx2}
  \frac{d \ln V}{d\varpi} &=& \frac{d \ln V/d\phi}{d\varpi/d\phi} = \frac{ \frac{\bar{\alpha}}{\sqrt{3}}\varpi - \left(\frac{\sqrt{3}\lambda}{\varpi - \sqrt{3}} + \frac{\sqrt{3}\lambda}{\varpi + \sqrt{3}}\right)}{\frac{\alpha}{2\sqrt{3}}\varpi^2  - \frac{\sqrt{3}(\alpha + 2\lambda)}{2}}
\end{eqnarray}
and
\begin{eqnarray}\label{zy2}
  \dot{\phi} &=& \varpi \sqrt{\frac{2V}{3 - \varpi^2}}, \quad H = \frac{1}{M_p} \sqrt{\frac{V}{3 - \varpi^2}}
\end{eqnarray}
where \( \varpi = \sqrt{3}/y \).

Inflationary evolution may be characterized by the Hubble-flow parameters \cite{Hoffman:2000ue,Schwarz:2001vv},
\begin{eqnarray}
  \epsilon_1 &\equiv& -\frac{\dot{H}}{H^2} = \varpi^2 \\
  \epsilon_2 &\equiv& \frac{\dot{\epsilon}_1}{H\epsilon_1} = \frac{2\alpha}{\sqrt{6}}\varpi^2  - \sqrt{6}(\alpha + 2\lambda) \\
  \epsilon_3 &\equiv& \frac{\dot{\epsilon}_2}{H\epsilon_2} = \frac{2\alpha}{\sqrt{6}}\varpi^2
\end{eqnarray}

Having excluded this degenerate limit, we now distinguish two principal nondegenerate cases, the second of which contains three subcases.

For \( \alpha = 0 \), Eq.~\ref{hx1} yields
\begin{eqnarray}
  \varpi &=& -\frac{\sqrt{3}\lambda}{M_p}\phi + C_1
\end{eqnarray}
where \( C_1 \) is an integration constant. Substituting this result into Eq.~\ref{hx2} gives
\begin{eqnarray}
  V(\varpi) &=& V_0(3 - \varpi^2)\exp\left[\frac{\sqrt{6}\varpi^2}{6\lambda}\right]
\end{eqnarray}
where \( V_0 \) is a positive constant.

For the second case, we take \( \alpha \neq 0 \) and \( \frac{\alpha}{2\sqrt{3}}\varpi^2  - \frac{\sqrt{3}(\alpha + 2\lambda)}{2} = c_0[\varpi^2 + b] \), or equivalently \( \alpha = 2\sqrt{3}c_0, \,\, \lambda = -\frac{1}{\sqrt{3}}c_0 b - \sqrt{3}c_0 \). Equation~\ref{hx1} then gives
\begin{eqnarray}
  \varpi(\phi) &=& \left\{
  \begin{array}{ll}
  \frac{1}{C_1 - c_0\frac{\phi}{M_p}}, & \text{if } b = 0, \\
  \sqrt{b} \tan\left( c_0\sqrt{b}\frac{\phi}{M_p} + C_1 \right), & \text{if } b > 0, \\
  \sqrt{-b} \frac{1 + C_1 \exp[2c_0\sqrt{-b}\frac{\phi}{M_p}]}{1 - C_1 \exp[2c_0\sqrt{-b}\frac{\phi}{M_p}]}, & \text{if } b < 0.
  \end{array}
  \right.
\end{eqnarray}
where \(C_1\) is an integration constant. The corresponding potential is
\begin{eqnarray}
  V(\varpi) &=& V_0|b + \varpi^2|^{-\frac{\sqrt{2}}{2c_0}}(3 - \varpi^2)
\end{eqnarray}

In the following we take $V_0>0$ and work on monotonic positive-$\varpi$
branches with $0<\varpi^2<1$ during accelerated expansion.  When
$b<0$, the singular point $\varpi^2=-b$ separates distinct branches
and is never crossed.  All integrals and numerical scans are therefore
restricted to one connected interval on which $\dot\phi\neq0$ and
$\varpi^2+b$ has a fixed sign.

This family contains previously studied solutions, including constant-roll inflation \cite{Motohashi:2014ppa,Yi:2017mxs}, for which the Hubble slow-roll parameter \( \eta_H \) is constant. From the definition
\[
  \eta_H = -\frac{\ddot{\phi}}{H\dot{\phi}} = 3 + \frac{V'}{H\dot{\phi}},
\]
and using Eq.~\ref{zy2}, one obtains
\begin{eqnarray}
  \frac{d \ln V}{d \phi} = f(y) &=& \frac{1}{M_p}\left[\frac{\lambda}{y - 1} + \frac{\lambda}{y + 1}\right]
\end{eqnarray}
where \( \lambda = \frac{\eta_H - 3}{\sqrt{6}} \).

Constant-roll inflation therefore corresponds to the special case
\(\bar{\alpha}=0\), with
\(\lambda=(\eta_H-3)/\sqrt{6}\).

\section{Physical Properties of the Exact Solutions}\label{sec:physical significance}
\subsection{Local attractor behavior of the rolling solutions}
\label{subsec:local-attractor}

We now investigate whether the exact rolling backgrounds are
local attractors in phase space. We assume an expanding
background for which the inflaton is monotonic and
$\dot{\phi}\neq0$ on the interval under consideration. Nearby
trajectories are compared at the same value of $\phi$. The
asymptotic limits at which $\dot{\phi}\rightarrow0$ are
understood as limits approached from within the rolling regime.

Following the linearized Hamilton--Jacobi attractor analysis of
Ref.~\cite{Liddle:1994dx}, we write
\[
H(\phi)=H_0(\phi)+\delta H(\phi),
\]
where $H_0(\phi)$ is an exact background solution and
$\delta H$ represents a nearby solution associated with the
same potential. The Hamilton--Jacobi equations are
\[
\dot{\phi}=-2M_p^2 H_{,\phi},
\qquad
V(\phi)=3M_p^2H^2-2M_p^4H_{,\phi}^2.
\]
Linearizing the second equation around $H_0(\phi)$ gives
\[
H_{0,\phi}\,\delta H_{,\phi}
=
\frac{3}{2M_p^2}H_0\,\delta H.
\]

Introducing the forward e-folding number
\[
\mathcal{N}\equiv\ln a,
\qquad
d\mathcal{N}=Hdt,
\]
and using
\[
\frac{d\phi}{d\mathcal{N}}
=
\frac{\dot{\phi}}{H}
=
-2M_p^2\frac{H_{0,\phi}}{H_0},
\]
we obtain
\[
\frac{d\ln|\delta H|}{d\mathcal{N}}=-3.
\]
Thus, $\delta H\propto a^{-3}$. This result follows directly
from the linearized Hamilton--Jacobi equation and does not rely
on the slow-roll approximation.

To characterize convergence of the field velocity toward the
rolling solution, we introduce the relative velocity
perturbation
\[
\Delta
\equiv
\frac{\delta\dot{\phi}}{\dot{\phi}}
=
\frac{\delta H_{,\phi}}{H_{0,\phi}}.
\]
Using the linearized Hamilton--Jacobi equation together with
\[
\epsilon_1
=
2M_p^2
\left(
\frac{H_{0,\phi}}{H_0}
\right)^2,
\]
we find
\[
\Delta
=
\frac{3}{\epsilon_1}\frac{\delta H}{H_0}.
\]
Since
\[
\frac{d\ln H_0}{d\mathcal{N}}=-\epsilon_1,
\qquad
\frac{d\ln\epsilon_1}{d\mathcal{N}}=\epsilon_2,
\]
the relative perturbation satisfies
\[
\frac{d\ln|\Delta|}{d\mathcal{N}}
=
\epsilon_1-\epsilon_2-3.
\]
Equivalently,
\[
\frac{d\Delta}{d\mathcal{N}}
=
\mathcal{A}\Delta,
\qquad
\mathcal{A}
\equiv
\epsilon_1-\epsilon_2-3.
\]
Its solution is
\[
\Delta(\mathcal{N})
=
\Delta_i
\exp\left[
\int_{\mathcal{N}_i}^{\mathcal{N}}
\mathcal{A}(\widetilde{\mathcal{N}})
d\widetilde{\mathcal{N}}
\right].
\]
Therefore,
\[
\mathcal{A}<0
\]
is a linear local criterion for attraction. More generally,
contraction of the relative perturbation over a finite interval
requires
\[
\int_{\mathcal{N}_i}^{\mathcal{N}}
\mathcal{A}\,
d\widetilde{\mathcal{N}}<0,
\]
whereas an asymptotic attractor requires this integral to tend
to $-\infty$. The condition $\mathcal{A}<0$ at a single point
only demonstrates local contraction and does not constitute a
proof of global nonlinear stability.

For the first family, $\alpha=0$, the Hubble-flow parameters are
\[
\epsilon_1=\varpi^2,
\qquad
\epsilon_2=-2\sqrt{6}\lambda.
\]
Consequently,
\[
\mathcal{A}(\varpi)
=
\varpi^2+2\sqrt{6}\lambda-3.
\]
At a finite point $\varpi=\varpi_*$, the local-attractor
condition is
\[
\lambda
<
\frac{3-\varpi_*^2}{2\sqrt{6}}.
\]
The direction of the background evolution follows from
Eq.~\eqref{hx1}:
\[
\dot{\varpi}
=
\frac{d\varpi}{d\phi}\dot{\phi}
=
-\frac{\sqrt{3}\lambda}{M_p}\,
\varpi
\sqrt{\frac{2V}{3-\varpi^2}}.
\]
For $\varpi>0$, the solution approaches $\varpi=0$ in forward
time when $\lambda>0$. In this asymptotic limit,
\[
\mathcal{A}
\longrightarrow
2\sqrt{6}\lambda-3,
\]
and the future-directed rolling solution is locally attracting
when
\[
0<\lambda<\frac{\sqrt{6}}{4}.
\]
For a finite rolling interval, the stronger pointwise condition
$\mathcal{A}(\varpi)<0$ must be checked throughout the interval.
For $\lambda<0$, $\varpi$ increases in forward time and the
solution evolves away from $\varpi=0$. This branch is therefore
not part of the preceding asymptotic classification, but it can
still be a linear local attractor over a finite inflationary
interval. The $\alpha=0$ parameter scans presented below use this
finite $\lambda<0$ branch and impose $\mathcal{A}<0$ pointwise.

For the second family, $\alpha\neq0$, the parametrization
introduced above gives
\[
\frac{d\varpi}{d\phi}
=
\frac{c_0}{M_p}(\varpi^2+b),
\]
and hence
\[
\epsilon_1=\varpi^2,
\qquad
\epsilon_2=2\sqrt{2}c_0(\varpi^2+b).
\]
The local-attractor coefficient is therefore
\[
\mathcal{A}(\varpi)
=
(1-2\sqrt{2}c_0)\varpi^2
-2\sqrt{2}bc_0-3.
\]
The direction of the background evolution is determined
directly by
\[
\dot{\varpi}
=
\frac{c_0}{M_p}\,
\varpi(\varpi^2+b)
\sqrt{\frac{2V}{3-\varpi^2}}.
\]

For the local analysis it is useful to introduce the combination
\[
u\equiv c_0(\epsilon_1+b),
\qquad
\epsilon_2=2\sqrt{2}u.
\]
At a sampled point the two quantities relevant for the rolling
solution are therefore
\[
\mathcal{A}
=\epsilon_1-2\sqrt{2}u-3,
\qquad
\frac{d\epsilon_1}{d\mathcal{N}}
=2\sqrt{2}u\epsilon_1.
\]
The observationally relevant red-spectrum region found below has
$u>0$ and $\epsilon_{1*}\ll1$. Consequently,
$\mathcal{A}_*<0$ throughout that region and $\epsilon_1$
increases in forward time.

The sign of $b$ and the position relative to the root
$\epsilon_1=-b$ divide this region into four dynamically distinct
branches:
\[
\begin{array}{lll}
{\rm (I)}
& b=0, & c_0>0,\\[1mm]
{\rm (II)}
& b>0, & c_0>0,\\[1mm]
{\rm (III)}
& b<0,\quad 0<\epsilon_1<-b<1, & c_0<0,\\[1mm]
{\rm (IV)}
& -1<b<0,\quad -b<\epsilon_1<1, & c_0>0.
\end{array}
\]
Branches (I), (II), and (IV) evolve toward the finite boundary
$\epsilon_1=1$, whereas branch (III) approaches the stable rolling
limit $\epsilon_1=-b<1$.  The numerical scan imposes the exact
pointwise condition
\[
(1-2\sqrt{2}c_0)\epsilon_{1*}
-2\sqrt{2}bc_0-3<0
\]
on every plotted point.  This is a local statement along the
rolling trajectory; it is not a claim of global nonlinear
stability.

Finally, constant-roll inflation corresponds to
\[
c_0=\frac{1}{\sqrt{2}},
\qquad
\eta_H=-b.
\]
In this case,
\[
\mathcal{A}
=
2\eta_H-3-\epsilon_1.
\]
Thus, at a finite point, the local-attractor condition is
\[
\eta_H<\frac{3+\epsilon_1}{2},
\]
which reduces in the quasi-de Sitter limit
$\epsilon_1\rightarrow0$ to
\[
\eta_H<\frac{3}{2}.
\]
This agrees with the constant-roll attractor condition obtained
from the two independent background modes and from numerical
phase-space analyses
\cite{Motohashi:2014ppa,Lin:2019fcz,
Motohashi:2019rhu,Motohashi:2025qgd}.

For the future-directed constant-roll branch approaching
$\varpi=0$, one has $b<0$ and therefore $\eta_H>0$. Combining
this fact with the local-attractor condition gives
\[
0<\eta_H<\frac{3}{2}.
\]
By contrast, the $b>0$ branch has $\eta_H<0$ and evolves away
from $\varpi=0$. It can still be a local rolling attractor, but
it is not part of the asymptotic $\varpi\rightarrow0$
classification used below.

\subsection{Number of e-folds}
Since $H = \dot{a}/a$, the expansion between the onset and the end of inflation can be characterized by the number of e-folds, $N$,
\begin{eqnarray}
 N &\equiv& \ln\frac{a(\phi_{\text{end}})}{a(\phi_{\text{start}})} = \int_{t_s}^{t_e} H dt = \int_{\phi_s}^{\phi_e} \frac{H}{\dot{\phi}} d\phi = \int_{\varpi_s}^{\varpi_e} \frac{1}{\sqrt{2}\varpi M_p} \frac{d\phi}{d\varpi} d\varpi \nonumber \\
 && = \int_{\varpi_s}^{\varpi_e} \frac{1}{\sqrt{2}\varpi} \frac{1}{\frac{\alpha}{2\sqrt{3}}\varpi^2 - \frac{\sqrt{3}(\alpha + 2\lambda)}{2}} d\varpi
\end{eqnarray}
As in the construction of $y$, $\phi$, and the potential $V$, this integral must be evaluated separately for each case.

For the first case, $\alpha = 0$, one finds
\begin{eqnarray}
 N &=& -\frac{1}{\sqrt{6}\lambda} \ln|\frac{\varpi_e}{\varpi_s}|, \,\,\,\, a = a_0 \varpi^{-\frac{1}{\sqrt{6}\lambda}}
\end{eqnarray}
On the finite $\lambda<0$ branch, choosing the end of accelerated
expansion at $\epsilon_{1e}=\varpi_e^2=1$ gives
\[
N_*=\frac{1}{2\sqrt{6}|\lambda|}
\ln\left(\frac{1}{\epsilon_{1*}}\right).
\]
This relation is not imposed in the $(n_s,r)$ scans below; it is
quoted to distinguish compatibility with the two-point observables
from the additional benchmark requirement that the pivot scale leave
the horizon approximately $50$--$60$ e-folds before the end of
inflation.  The precise interval depends on reheating and the
post-inflationary expansion history and is not itself a direct
observational bound \cite{Liddle:2003as}.

For the second case, $\alpha \neq 0, \frac{\alpha}{2\sqrt{3}} \varpi^2 - \frac{\sqrt{3}(\alpha + 2\lambda)}{2} = c_0[\varpi^2 + b]$. When $b = 0$,
\begin{eqnarray}
 N &=& -\frac{1}{2\sqrt{2}c_0} \left( \varpi_e^{-2} - \varpi_s^{-2} \right), \,\,\,\, a = a_0 \exp\left[-\frac{1}{2\sqrt{2}c_0 \varpi^2}\right]
\end{eqnarray}
whereas for $b \neq 0$,
\begin{eqnarray}
 N &=& \frac{1}{2\sqrt{2}c_0 b} \ln \left[ \frac{\varpi_e^2}{\varpi_s^2} \left| \frac{\varpi_s^2 + b}{\varpi_e^2 + b} \right| \right], \,\,\,\, a = a_0 \left| \frac{\varpi^2}{\varpi^2 + b} \right|^{\frac{1}{2\sqrt{2}c_0 b}}
\end{eqnarray}
For a branch that reaches $\epsilon_{1e}=1$, these expressions give
\[
N_*=\frac{1}{2\sqrt{2}c_0}
\left(\frac{1}{\epsilon_{1*}}-1\right)
\qquad (b=0),
\]
and
\[
N_*=
\frac{1}{2\sqrt{2}bc_0}
\ln\left|
\frac{\epsilon_{1*}+b}
{\epsilon_{1*}(1+b)}
\right|
\qquad (b\neq0).
\]
The condition for accelerated expansion follows from
\begin{eqnarray}
 \frac{\ddot{a}}{a} &=& H^2 + \dot{H} = \frac{V}{M_p^2 (3 - \varpi^2)}(1 - \varpi^2)
\end{eqnarray}

The above expressions characterize the amount of expansion generated
along the exact rolling backgrounds.  We first identify the regions
compatible with the local-attractor, superhorizon, and observational
conditions and then evaluate $N_*$ on those regions.  Branches (I),
(II), and (IV) reach $\epsilon_1=1$.  Branch (II) admits a nonempty
intersection with $50\leq N_*<60$ after $b$ is allowed to vary; a
representative choice is $b=0.003$.  Branches (I) and (IV), however,
give $N_*<50$ throughout their joint-CMB-compatible portions.
Branch (III) approaches the fixed rolling value
$\epsilon_1=-b<1$ and never reaches $\epsilon_1=1$, so an added exit
sector would be required; hybrid inflation provides a standard example
of such a mechanism \cite{Linde:1993cn}.  We distinguish this
behavior from a trajectory with $N_*>60$: the latter does reach
$\epsilon_1=1$ but lies outside the adopted pivot-to-end e-fold window.
Here ``exit'' only means the crossing of $\epsilon_1=1$, i.e. the end
of accelerated expansion; the subsequent reheating transition is not
modeled.

\subsection{Superhorizon evolution}
Outside the Hubble radius, the curvature perturbation $\zeta$ can
in general evolve.  Its long-wavelength solution is
\begin{eqnarray}
 \zeta &=& A_k + B_k \int^{\tau}
 \frac{d\bar{\tau}}{z^2(\bar{\tau})},
 \qquad z=\frac{a\dot{\phi}}{H}.
\end{eqnarray}
The first term is constant, while the second is the time-dependent
mode. Since $z^2=2a^2M_p^2\epsilon_1$ and
$d\tau/d\mathcal{N}=1/(aH)$, its local evolution obeys
\[
\frac{d\zeta_k}{d\mathcal{N}}
=\frac{B_k}{2M_p^2a^3H\epsilon_1},
\]
and hence
\[
\frac{d}{d\mathcal{N}}
\ln\left|
\frac{d\zeta_k}{d\mathcal{N}}
\right|
=-3+\epsilon_1-\epsilon_2
=\mathcal{A}.
\]
Therefore, the same pointwise condition
\[
\mathcal{A}=\epsilon_1-\epsilon_2-3<0
\]
that gives local contraction of the rolling background also makes
the derivative of the time-dependent superhorizon mode decrease
locally.  This is the superhorizon restriction imposed in the
parameter scans below.

For the finite $\alpha=0$, $\lambda<0$ branch, and for branches
(I), (II), and (IV) of the $\alpha\neq0$ family, the condition is
checked pointwise up to the finite reference boundary
$\epsilon_1=1$.  For branch (III), which approaches
$\epsilon_1=-b<1$ asymptotically, one has
$\mathcal{A}\rightarrow-b-3<0$; the corresponding
time-dependent mode decays toward that rolling limit.  No stronger
global convergence condition is imposed.  The finite-$N_*$ cut used
for branch (II) below is an additional background requirement and
does not modify this local superhorizon criterion.

For constant-roll inflation with $\alpha=\sqrt{6}$, or equivalently $c_0=1/\sqrt{2}$, the second Hubble slow-roll parameter is
\[
 \eta_H=\sqrt{6}\lambda+3=-b.
\]
The local rolling-attractor condition derived in Sec.~\ref{subsec:local-attractor} reduces in the quasi-de Sitter limit to $\eta_H<3/2$. For the future-directed constant-roll branch approaching $\varpi=0$, one has $b<0$ and hence $\eta_H=-b>0$. Combining these conditions gives
\[
 0<\eta_H<\frac{3}{2}.
\]
The same upper bound follows independently from the decay of the time-dependent superhorizon mode, in agreement with Refs.~\cite{Lin:2019fcz,Gao:2019sbz,Motohashi:2019rhu,Motohashi:2025qgd}.
\subsection{Scalar perturbations}
The scalar mode function obeys the Mukhanov-Sasaki equation \cite{Mukhanov:1985rz,Sasaki:1986hm},
\[
v_k'' + \left( k^2 - \frac{z''}{z} \right) v_k = 0
\]
where \( z = a \dot{\phi} / H \), and a prime denotes differentiation with respect to conformal time \( \tau \). When the effective index varies sufficiently slowly, we define the local scalar index \( \nu_R \) by
\[
Q_R\equiv\tau^2\frac{z''}{z}=\nu_R^2-\frac14,
\qquad n_s-1\simeq3-2\nu_R,
\]
and
\[
\frac{z''}{z} = a^2 H^2 \left( 2 - \epsilon_1 + \frac{3}{2} \epsilon_2 + \frac{1}{4} \epsilon_2^2 - \frac{1}{2} \epsilon_1 \epsilon_2 + \frac{1}{2} \epsilon_2 \epsilon_3 \right)
\]

To relate the background evolution to conformal time, one must also determine the time direction in which \( \varpi \) evolves,
\[
\tau = \int^{\varpi}_{\varpi_e} \frac{d\varpi'}{a(\varpi')\dot{\varpi}(\varpi')}
\]
where \( \varpi \) denotes the relevant background value.  The
additive constant in conformal time is fixed by setting $\tau=0$ at a
finite future endpoint $\varpi_e$, or by taking $\tau\to0$ at a future
asymptotic endpoint.  This convention is used consistently in the
local quantities $Q_R=\tau^2z''/z$ and $Q_T=\tau^2a''/a$. We again
consider the different cases separately.

For the first case, \( \alpha = 0 \), the future-directed
asymptotic branch with $\lambda>0$ approaches $\varpi_e=0$ and
\begin{eqnarray}
 \tau &=& -\frac{M_p}{\lambda a_0 \sqrt{6V_0}} \int^{\varpi}_0 \omega^{\frac{1}{\sqrt{6}\lambda} - 1} \exp\left( -\frac{\sqrt{6}}{12\lambda} \omega^2 \right) d\omega \nonumber\\
&=& \frac{M_p}{2 \lambda a_0 \sqrt{6V_0}} \left( \frac{\sqrt{6}}{12\lambda} \right)^{-\frac{\sqrt{6}}{12\lambda}} \left[\Gamma \left( \frac{\sqrt{6}}{12\lambda}, \frac{\sqrt{6}}{12\lambda} \varpi^2 \right)-\Gamma \left( \frac{\sqrt{6}}{12\lambda}, 0 \right)\right]
\end{eqnarray}
where $\Gamma(s, x) = \int_x^{\infty} t^{s-1} e^{-t} \, dt$ is the upper incomplete gamma function.
For the finite $\lambda<0$ branch, $\varpi$ increases in forward
time and the preceding expression with lower endpoint
$\varpi_e=0$ is not applicable. Instead, for a finite future
endpoint $\varpi_e>\varpi$ one must use the manifestly real
integral
\[
\tau(\varpi)
=
-\frac{M_p}{\lambda a_0\sqrt{6V_0}}
\int_{\varpi_e}^{\varpi}
\omega^{\frac{1}{\sqrt{6}\lambda}-1}
\exp\left(-\frac{\sqrt{6}}{12\lambda}\omega^2\right)d\omega,
\qquad \lambda<0,
\]
with $\tau(\varpi_e)=0$. In Fig.~\ref{tu1} we take
$\varpi_e=1$, or equivalently $\epsilon_{1e}=1$. We evaluate this
real integral numerically rather than analytically continuing the
incomplete-gamma expression through $\Gamma(s,0)$ for $s<0$.
\[
\frac{z''}{z} = \frac{a_0^2}{M_p^2} V_0 \exp\left( \frac{\sqrt{6}}{6\lambda} \varpi^2 \right) \left[ 2 - 3\sqrt{6} \lambda + 6 \lambda^2 + \left( -1 + \sqrt{6} \lambda \right) \varpi^2 \right] (\varpi^2)^{-\frac{1}{\sqrt{6}\lambda}}
\]
The asymptotic expression with endpoint $\varpi_e=0$ applies for
$0<\lambda<\sqrt{6}/4$ and $\varpi>0$, whereas the finite-endpoint
integral above is used for the negative-$\lambda$ branch.
The remaining cases can be treated in the same manner.

For the second case, \( c_0 \neq 0 \), the result depends on the sign of $b$.

When \( b = 0 \), the real finite-endpoint expression is
\[
\tau(\varpi)=\frac{M_p}{c_0 a_0 \sqrt{2V_0}}
\int^{\varpi}_{\varpi_e}
\omega^{\frac{\sqrt{2}}{2 c_0} -3}
\exp\left(\frac{1}{2\sqrt{2}c_0\omega^2}\right)d\omega .
\]
For the $c_0>0$ branch used in Fig.~\ref{tu2}(I), we take
$\varpi_e=1$.  The lower endpoint cannot be set to zero on this branch
because the exponential diverges as $\omega\to0^+$.  For the
$c_0<0$ asymptotic branch, on the other hand, one may take
$\varpi_e=0$ and write
\[
\tau=\frac{M_p}{2 c_0 a_0 \sqrt{2V_0}}
\left(-2\sqrt{2}c_0\right)^{1-\frac{\sqrt{2}}{4c_0}}
\Gamma\left(1-\frac{\sqrt{2}}{4c_0},
-\frac{1}{2\sqrt{2}c_0\varpi^2}\right).
\]
\[
\frac{z''}{z} = \frac{a_0^2}{M_p^2} V_0 |\varpi|^{-\frac{\sqrt{2}}{c_0 }} \exp(-\frac{1}{\sqrt{2}c_0\varpi^2})\left[ 2 + (-1+3\sqrt{2} c_0) \varpi^2  + (6c_0^2 -\sqrt{2}c_0 ) \varpi^4 \right]
\]

When \( b \neq 0 \),
\[
\tau = \frac{M_p}{c_0 a_0 \sqrt{2V_0}}
\int^{\varpi}_{\varpi_e}
\operatorname{sgn}(\omega^2+b)
\omega^{-\frac{\sqrt{2}}{2b c_0} - 1}
|\omega^2 + b|^{\frac{\sqrt{2}}{4c_0}
+ \frac{\sqrt{2}}{4b c_0} - 1} d\omega .
\]
\[
\frac{z''}{z} = \frac{a_0^2}{M_p^2} V_0 |\varpi^2|^{\frac{1}{\sqrt{2}c_0 b}} |\varpi^2 + b|^{-\frac{1}{\sqrt{2}c_0}(1 + \frac{1}{b})}
\]
\[
\times \left[ 2 - \varpi^2+ 2 c_0^2 (\varpi^2 + b)^2 + (4c_0^2 - \sqrt{2} c_0) \varpi^2 (\varpi^2 + b) + 3\sqrt{2} c_0 (\varpi^2 + b)  \right]
\]
This absolute-value form is real on every connected branch and is the
form used numerically.  For branches (II) and (IV) in
Fig.~\ref{tu2}, $\varpi_e=1$ and $\tau(1)=0$.  Branch (III) has the
future asymptotic endpoint $\varpi_e=\sqrt{-b}$.  The following
hypergeometric expressions are useful closed forms for the indicated
asymptotic branches.
For $b>0$, $c_0<0$, and $\varpi>0$, with $-3\sqrt{2}/4<bc_0<0$, we choose the future asymptotic reference value $\varpi_f=0$ and obtain
\[
\tau = - \frac{M_p}{a_0 \sqrt{V_0}} b^{\frac{1 + b}{2 \sqrt{2} b c_0}} \varpi^{-\frac{1}{\sqrt{2} b c_0}} {}_2 F_1 \left( -\frac{1}{2 \sqrt{2} b c_0}, 1 - \frac{\sqrt{2} + \sqrt{2} b}{4 b c_0}; 1 - \frac{1}{2 \sqrt{2} b c_0}; -\frac{\varpi^2}{b} \right)
\]
where \({}_2F_1\) is the Gauss hypergeometric function. For $b<0$, $c_0>0$, $0<\varpi^2<-b$, and $-3\sqrt{2}/4<bc_0<0$, the future asymptotic value is again $\varpi_f=0$, and
\[
\tau = - \frac{M_p}{a_0 \sqrt{V_0}} |b|^{\frac{1 + b}{2 \sqrt{2} b c_0}} \varpi^{-\frac{1}{\sqrt{2} b c_0}} {}_2 F_1 \left( -\frac{1}{2 \sqrt{2} b c_0}, 1 - \frac{ \sqrt{2} + \sqrt{2} b}{4 b c_0}; 1 - \frac{1}{2 \sqrt{2} b c_0}; -\frac{\varpi^2}{b} \right)
\]
For the future-directed asymptotic branch with $b<0$, $c_0<0$, and $0<\varpi^2<-b<1$, one has $\varpi_f=\sqrt{-b}$, and
\[
|\tau| = \pm \frac{ M_p}{1 + b} \frac{|b|^{-\frac{1}{2 \sqrt{2} b c_0}}}{ a_0 \sqrt{V_0}} |\varpi^2 + b|^{\frac{1 + b}{2 \sqrt{2} b c_0}} {}_2 F_1 \left( 1 + \frac{1}{2 \sqrt{2} b c_0}, \frac{1 + b}{2 \sqrt{2} b c_0}; 1 + \frac{1 + b}{2 \sqrt{2} b c_0}; \frac{\varpi^2 + b}{b} \right)
\]
\subsection{Tensor perturbations}
The tensor mode function \( u_k \) satisfies
\[
u_k'' + \left( k^2 - \frac{a''}{a} \right) u_k = 0,
\qquad
Q_T\equiv\tau^2\frac{a''}{a}=\mu_T^2-\frac14
=\tau^2a^2H^2(2-\varpi^2),
\]
and, within the same local-index approximation, the tensor-to-scalar ratio is
\[
n_T\simeq3-2\mu_T,
\]
and, when the scalar and tensor spectra are evaluated at the common
Hubble-crossing condition $k=aH$,
\[
r \simeq \frac{\mathcal{P_T}}{\mathcal{P_R}}
= 16 \varpi^2 2^{2(\mu_T - \nu_R)}
\frac{\Gamma^2(\mu_T)}{\Gamma^2(\nu_R)}
|aH\tau|^{2(\nu_R-\mu_T)}.
\]
If the local modes are instead matched at $-k\tau=1$, the final
factor is unity.  All numerical results below use $k=aH$ and retain
the $|aH\tau|$ factor.

\subsection{Consistency of the local-index approximation}
The preceding expressions for $n_s$ and $r$ use the Hankel-function
form of the scalar and tensor modes with $\nu_R$ and $\mu_T$ treated
as locally constant.  This assumption can be tested directly without
introducing a normalization-dependent derivative.  We define
\[
\Delta_R\equiv\frac{d\ln Q_R}{d\ln|\tau|},
\qquad
\Delta_T\equiv\frac{d\ln Q_T}{d\ln|\tau|}.
\]
The local-index approximation requires
$|\Delta_R|\ll1$ and $|\Delta_T|\ll1$.  The corresponding derivatives
with respect to conformal time are
\[
\frac{dQ_R}{d\tau}=\frac{Q_R}{\tau}\Delta_R,
\qquad
\frac{dQ_T}{d\tau}=\frac{Q_T}{\tau}\Delta_T.
\]
Thus the small quantities to be compared are the dimensionless
logarithmic rates rather than the dimensionful derivatives
$dQ_R/d\tau$ and $dQ_T/d\tau$ themselves.

To express these rates in terms of the exact rolling variable, let
\[
\epsilon\equiv\epsilon_1=\varpi^2,
\qquad x(\varpi)\equiv aH\tau,
\]
and write
\[
Q_R=x^2F_R,
\qquad Q_T=x^2F_T,
\]
where
\[
F_R=2-\epsilon+\frac32\epsilon_2+\frac14\epsilon_2^2
-\frac12\epsilon\epsilon_2+\frac12\epsilon_2\epsilon_3,
\qquad F_T=2-\epsilon.
\]
The exact background identities
\[
\frac{dx}{d\mathcal N}=(1-\epsilon)x+1,
\qquad
\frac{d\epsilon}{d\mathcal N}=\epsilon\epsilon_2,
\qquad
\frac{d\ln|\tau|}{d\mathcal N}=\frac1x
\]
then give
\begin{align}
\Delta_R
&=2+x\left[2(1-\epsilon)
+\frac{d\ln F_R}{d\mathcal N}\right],\label{DeltaR}\\
\Delta_T
&=2+x\left[2(1-\epsilon)
-\frac{\epsilon\epsilon_2}{2-\epsilon}\right].\label{DeltaT}
\end{align}
For the exact solutions considered here, $F_R$ is a function of
$\epsilon=\varpi^2$, so that
\[
\frac{d\ln F_R}{d\mathcal N}
=\frac{\epsilon\epsilon_2}{F_R}\frac{dF_R}{d\epsilon}.
\]
Equations~\eqref{DeltaR} and \eqref{DeltaT}, together with the exact
$\tau(\varpi)$ obtained above, therefore express both consistency
measures entirely as functions of $\varpi$.

For $\alpha=0$ the required functions reduce to
\begin{align}
F_R={}&2-3\sqrt6\lambda+6\lambda^2
+(-1+\sqrt6\lambda)\epsilon,\\
\frac{d\epsilon}{d\mathcal N}={}&-2\sqrt6\lambda\epsilon,
\qquad
\frac{dF_R}{d\epsilon}=-1+\sqrt6\lambda.
\end{align}
For $\alpha\neq0$ one instead has
\begin{align}
F_R={}&2-\epsilon+2c_0^2(\epsilon+b)^2
+(4c_0^2-\sqrt2c_0)\epsilon(\epsilon+b)
+3\sqrt2c_0(\epsilon+b),\\
\frac{d\epsilon}{d\mathcal N}={}&
2\sqrt2c_0\epsilon(\epsilon+b),\\
\frac{dF_R}{d\epsilon}={}&-1+4c_0^2(\epsilon+b)
+(4c_0^2-\sqrt2c_0)(2\epsilon+b)+3\sqrt2c_0.
\end{align}

\section{Numerical Results and Discussion}
Our primary observational reference is the recent CMB-only analysis of
Planck, SPT, ACT, and BICEP/Keck data, which gives
$n_s=0.9682\pm0.0032$ and $r_{0.05}<0.034$ at 95\% C.L.
\cite{Balkenhol:2025}.  We use the public 68\% and 95\% joint
$(n_s,r)$ contour lines released with that analysis, at
$k_*=0.05\,\mathrm{Mpc}^{-1}$, both for the shaded regions in the
figures and for the numerical membership test.  Thus every interval
quoted below is defined by the point lying inside the adopted 95\%
joint CMB contour, rather than by combining marginal one-dimensional
cuts.  The same analysis finds an upward shift to
$n_s=0.9728\pm0.0029$ when DESI BAO data are added, but notes a
marginal tension between the CMB and BAO data; we therefore keep the
CMB-only contour as the conservative primary comparison.  The earlier
Planck 2018 and BK18 results \cite{Planck2018,BK2021} are retained only
as historical reference points.

The scalar amplitude fixes the otherwise free normalization $V_0$.
Within the local-index treatment,
\[
\mathcal P_{\mathcal R}(k_*)\simeq
\frac{H_*^2}{8\pi^2M_p^2\epsilon_{1*}}
2^{2\nu_R-3}
\left[\frac{\Gamma(\nu_R)}{\Gamma(3/2)}\right]^2
|a_*H_*\tau_*|^{1-2\nu_R}.
\]
For the representative branch-(II) point used below, imposing
$A_s=2.1\times10^{-9}$ gives
\[
\frac{H_*}{M_p}\simeq1.27\times10^{-5},\qquad
\frac{V_0}{M_p^4}\simeq4.39\times10^{-11},\qquad
\frac{V_*^{1/4}}{M_p}\simeq4.69\times10^{-3}.
\]
Using $M_p=2.435\times10^{18}\,\mathrm{GeV}$, the last value is
$V_*^{1/4}\simeq1.14\times10^{16}\,\mathrm{GeV}$.  Hence the amplitude
can be matched by fixing $V_0$ without changing the dimensionless
$(n_s,r)$ trajectories.

The Hubble slow-roll parameters are \cite{Copeland:1993jj,Liddle:1994dx}
\begin{eqnarray}
  \epsilon_H &=& -\frac{\dot{H}}{H^2},\,\,\,\eta_H = -\frac{\ddot{\phi}}{H \dot{\phi}}
\end{eqnarray}
which for the present solutions become
\begin{eqnarray}
  \epsilon_H &=&\varpi^2,\,\,\,\eta_H = \frac{6 - \sqrt{6} \alpha}{6} \varpi^2 +  \frac{\sqrt{6}}{2} \left( \alpha + 2 \lambda \right)
\end{eqnarray}
For \( \alpha= \sqrt{6} \), one obtains the constant value \( \eta_H  =3+\sqrt{6}\lambda \), as in the constant-roll models of Refs.~\cite{Motohashi:2014ppa,Yi:2017mxs}.
Finally, the Hubble-flow parameters introduced above take the explicit form
\begin{eqnarray}
  \epsilon_1 &=&  \varpi^2
 , \,\,\,\epsilon_2 = \frac{2\alpha}{\sqrt{6}}\varpi^2  - \sqrt{6}(\alpha + 2\lambda)
\end{eqnarray}
For \( \alpha= 0 \), the second Hubble-flow parameter is constant, \( \epsilon_2  =-2\sqrt{6}\lambda \). We now discuss the allowed parameter space case by case.

\begin{figure}[t!]
\centering
\includegraphics[width=3.0in]{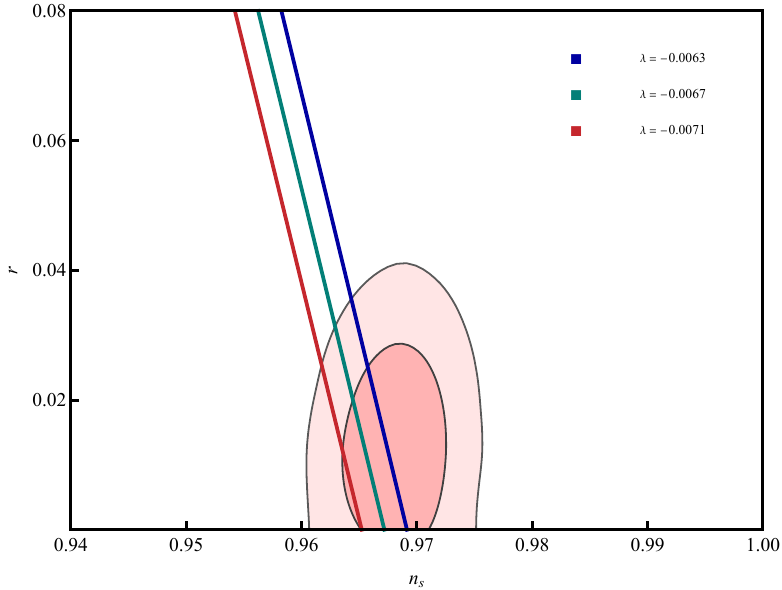}\hfill
\includegraphics[width=3.0in]{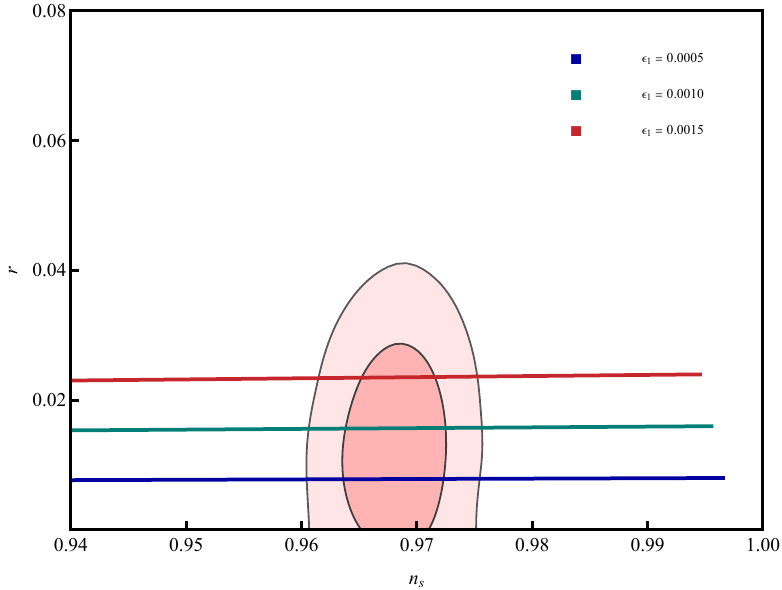}
\caption{Predictions in the $(n_s,r)$ plane for the finite
$\alpha=0$, $\lambda<0$ rolling branch. In the left panel,
$\lambda=-0.0063,-0.0067,-0.0071$ is fixed on each curve while
$\epsilon_{1*}$ is varied. In the right panel,
$\epsilon_{1*}=0.0005,0.0010,0.0015$ is fixed on each curve while
$\lambda$ is varied. We set $\epsilon_{1e}=1$ to define the finite
conformal-time endpoint and impose the linear local-attractor
condition $\mathcal{A}_*<0$ at every sampled point. The inner and outer
shaded regions are respectively the public 68\% and 95\% joint CMB
contours of Ref.~\cite{Balkenhol:2025}.}
\label{tu1}
\end{figure}

For the first case, $\alpha=0$, restricting attention to the
future-directed asymptotic branch with $\lambda>0$ and
$\varpi\rightarrow0$ yields no region compatible with both
$n_s$ and $r$. The finite $\lambda<0$ rolling branch is different:
$\varpi$ and $\epsilon_1$ increase in forward time, while the
pointwise coefficient
\[
\mathcal{A}_*=\epsilon_{1*}+2\sqrt{6}\lambda-3
\]
remains negative throughout the parameter ranges displayed in
Fig.~\ref{tu1}.

The two panels give complementary slices through the same finite
branch. They demonstrate that the constant-$\epsilon_2$ family has a
nonempty region inside the adopted 95\% joint CMB contour. For
example,
\[
\epsilon_{1*}=0.0010,\qquad \lambda=-0.0067
\]
gives
\[
n_s\simeq0.96505,\qquad r\simeq0.01562,
\qquad \epsilon_2\simeq0.03282,
\qquad \mathcal{A}_*\simeq-3.03182.
\]
However, no e-fold condition is imposed in Fig.~\ref{tu1}.
Using $\epsilon_{1e}=1$ in the background solution gives
$N_*\simeq210$ for this representative point, rather than
$50$--$60$.  The background does reach $\epsilon_1=1$, but the selected
pivot point lies outside the adopted e-fold window. Thus
Fig.~\ref{tu1} establishes compatibility with the joint $(n_s,r)$
contour and linear local attraction, but not a pivot-to-$\epsilon_1=1$
interval with $50\leq N_*<60$.

For the second case, $\alpha\neq0$, it is convenient to scan the
combination
\[
u_*=c_0(\epsilon_{1*}+b),
\qquad
\epsilon_{2*}=2\sqrt{2}u_*.
\]
\begin{figure*}[t!]
\centering
\includegraphics[width=0.48\textwidth]{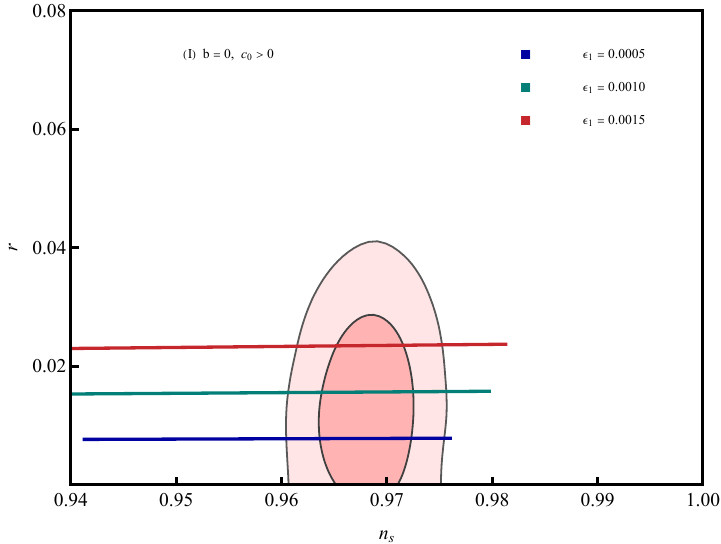}\hfill
\includegraphics[width=0.48\textwidth]{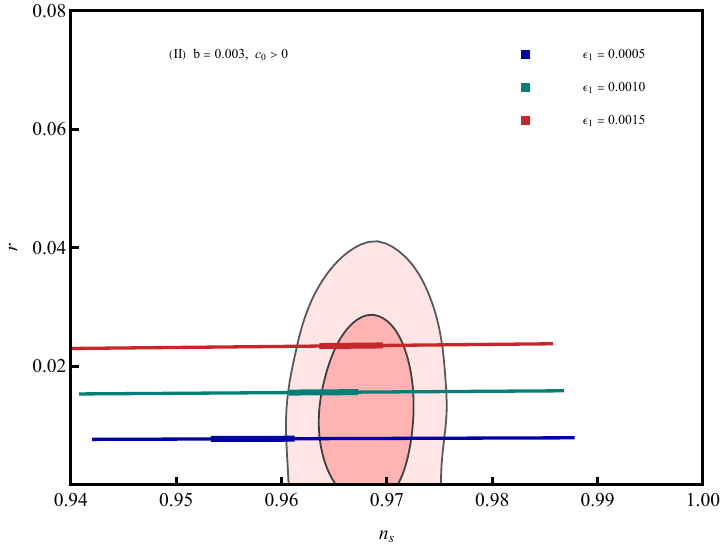}\\[2mm]
\includegraphics[width=0.48\textwidth]{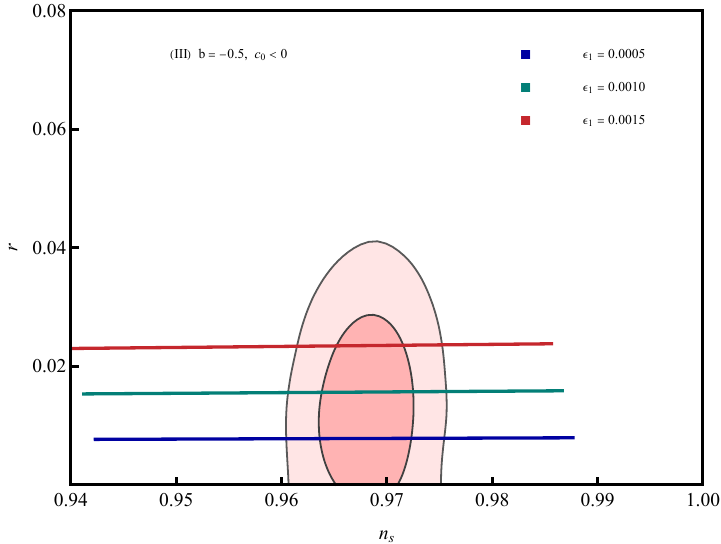}\hfill
\includegraphics[width=0.48\textwidth]{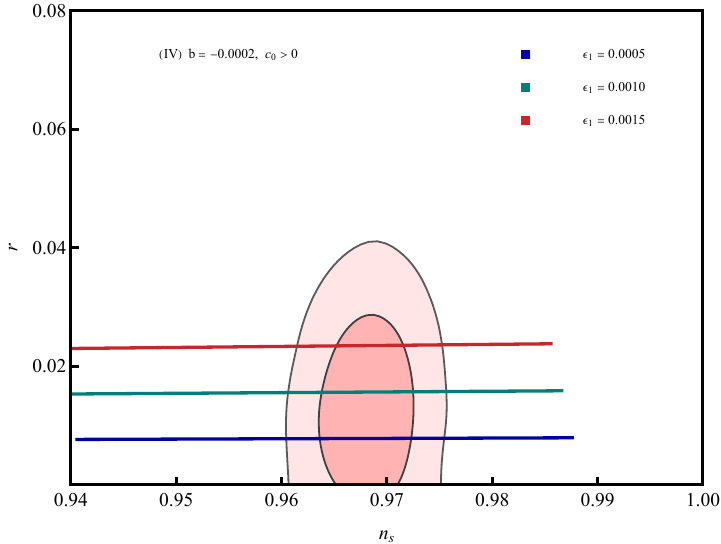}
\caption{Predictions in the $(n_s,r)$ plane for the four
$\alpha\neq0$ branches: (I) $b=0$, $c_0>0$;
(II) $b=0.003$, $c_0>0$;
(III) $b=-0.5$, $0<\epsilon_{1*}<-b$, $c_0<0$; and
(IV) $b=-0.0002$, $-b<\epsilon_{1*}<1$, $c_0>0$.
The three colors correspond to
$\epsilon_{1*}=0.0005$, $0.0010$, and $0.0015$, while $u_*$ is
varied. Every plotted point satisfies $\mathcal{A}_*<0$.  The complete
curves are shown in all panels; in panel (II), the thicker portions
also satisfy $50\leq N_*<60$.  The inner and outer shaded regions are
the public 68\% and 95\% joint CMB contours of
Ref.~\cite{Balkenhol:2025}. Branches (I), (II), and (IV) reach
$\epsilon_1=1$, whereas branch (III) approaches
$\epsilon_1=-b<1$.}
\label{tu2}
\end{figure*}

For the three displayed values
$\epsilon_{1*}=0.0005,0.0010,0.0015$, samples lying inside the adopted
95\% joint contour occur approximately in
\[
0.0078\lesssim u_*\lesssim0.0132,
\]
with the exact interval depending weakly on the dynamical branch and
on the e-fold requirement.

Figure~\ref{tu2} shows that all four branches contain segments inside
the adopted 95\% joint contour. In this region,
\[
\mathcal{A}_*
=\epsilon_{1*}-2\sqrt{2}u_*-3<0,
\]
so the local-attractor and local superhorizon restrictions are both
satisfied. The e-fold test, however, distinguishes the four cases.
Combining all three displayed values of $\epsilon_{1*}$, branch (I)
gives approximately $26.4\lesssim N_*\lesssim46.5$, while branch
(IV) gives $20.4\lesssim N_*\lesssim43.2$. Thus both branches reach
$\epsilon_1=1$ too early to satisfy the conventional lower bound.
Branch (III) never reaches $\epsilon_1=1$ because it approaches the
fixed rolling value $\epsilon_1=-b=0.5$.

Branch (II) is qualitatively different. For $b=0.003$ there is a
nonempty intersection of all the imposed conditions. A representative
point is
\[
\epsilon_{1*}=0.001,
\qquad b=0.003,
\qquad c_0=3,
\qquad u_*=0.012,
\]
for which
\[
N_*\simeq54.34,
\qquad n_s\simeq0.96383,
\qquad r\simeq0.01560,
\]
and
\[
\epsilon_{2*}\simeq0.03394,
\qquad \mathcal{A}_*\simeq-3.03294<0.
\]
This point is locally attracting, has a locally decreasing
time-dependent superhorizon mode, lies inside the joint 95\% CMB
contour, and reaches $\epsilon_1=1$ within the adopted e-fold window.

Representative intervals at $\epsilon_{1*}=0.001$ are listed in
Table~\ref{tab:secondcase-space}. For branch (II), the quoted interval
already includes the requirement $50\leq N_*<60$.
\begin{table}[htbp]
\centering
\caption{Approximate parameter intervals and e-fold outcomes for the
four $\alpha\neq0$ branches at $\epsilon_{1*}=0.001$.  The intervals
lie inside the adopted 95\% joint CMB contour; branch (II) additionally
satisfies $50\leq N_*<60$.}
\label{tab:secondcase-space}
\begin{tabular}{ccccc}
\specialrule{0.1em}{0pt}{1pt}
Branch & $b$ & Condition & Approximate $c_0$ interval & $N_*$ outcome \\
\specialrule{0.05em}{1pt}{1pt}
(I) & $0$ & $c_0>0$ & $7.85<c_0<13.00$ & $27.2$--$45.0$ \\
(II) & $0.003$ & $c_0>0$ & $2.725<c_0<3.250$ & $50.2$--$59.8$ \\
(III) & $-0.5$ & $\epsilon_{1*}<-b$, $c_0<0$ & $-0.0264<c_0<-0.0158$ & no $\epsilon_1=1$ crossing \\
(IV) & $-0.0002$ & $-b<\epsilon_{1*}$, $c_0>0$ & $9.81<c_0<16.19$ & $24.3$--$40.2$ \\
\specialrule{0.1em}{1pt}{0pt}
\end{tabular}
\end{table}
\FloatBarrier

We finally apply the local-index consistency test of
Eqs.~\eqref{DeltaR} and \eqref{DeltaT} to the same parameter grids.
Table~\ref{tab:index-consistency} reports the largest absolute rates
among the sampled points inside the adopted 95\% joint CMB contour.
For branch (II), the additional condition $50\leq N_*<60$ is
included.  The rates remain well below unity in every case.
\begin{table}[htbp]
\centering
\caption{Maximum dimensionless variation rates of the scalar and
tensor effective potentials over the parts of the displayed parameter
grids lying inside the adopted 95\% joint CMB contour.}
\label{tab:index-consistency}
\begin{tabular}{ccc}
\specialrule{0.1em}{0pt}{1pt}
Region & $\max|\Delta_R|$ & $\max|\Delta_T|$ \\
\specialrule{0.05em}{1pt}{1pt}
Fig.~\ref{tu1}, $\alpha=0$ & $1.10\times10^{-4}$ & $1.10\times10^{-4}$ \\
Fig.~\ref{tu2}(I) & $1.12\times10^{-3}$ & $8.67\times10^{-5}$ \\
Fig.~\ref{tu2}(II), $50\leq N_*<60$ & $3.49\times10^{-4}$ & $7.87\times10^{-5}$ \\
Fig.~\ref{tu2}(III) & $8.08\times10^{-5}$ & $8.34\times10^{-5}$ \\
Fig.~\ref{tu2}(IV) & $1.83\times10^{-3}$ & $8.75\times10^{-5}$ \\
\specialrule{0.1em}{1pt}{0pt}
\end{tabular}
\end{table}
\FloatBarrier
At the representative branch-(II) point quoted above, one finds
\[
Q_R\simeq2.05458,
\qquad \Delta_R\simeq-2.70\times10^{-4},
\qquad
Q_T\simeq2.00315,
\qquad \Delta_T\simeq-5.41\times10^{-5}.
\]
Thus both effective indices vary slowly near Hubble crossing, so the
necessary local adiabaticity condition for the Hankel-index treatment
is satisfied over the displayed parameter regions.

\subsection{Direct numerical mode validation}
We additionally validate the local-index calculation by evolving the
scalar and tensor modes directly on the exact backgrounds.  With
$\mathcal N=\ln(a/a_*)$, their equations can be written as
\begin{align}
v_{k,\mathcal NN}+(1-\epsilon_1)v_{k,\mathcal N}
 +\left[\left(\frac{k}{aH}\right)^2-F_R\right]v_k&=0,\\
u_{k,\mathcal NN}+(1-\epsilon_1)u_{k,\mathcal N}
 +\left[\left(\frac{k}{aH}\right)^2-F_T\right]u_k&=0.
\end{align}
We impose the Bunch--Davies vacuum \cite{Bunch:1978yq} when
$k/(aH)\simeq100$, integrate until the scalar and tensor spectra have
frozen, and determine the tilts by finite differences about $k_*$.  The
comparison in Table~\ref{tab:mode-validation} uses one representative
point from each displayed branch.
\begin{table}[htbp]
\centering
\caption{Local-index estimates and direct numerical mode results at
representative points.}
\label{tab:mode-validation}
\begin{tabular}{ccccc}
\specialrule{0.1em}{0pt}{1pt}
Region & $n_s^{\rm local}$ & $n_s^{\rm num}$ & $r^{\rm local}$ & $r^{\rm num}$ \\
\specialrule{0.05em}{1pt}{1pt}
$\alpha=0$ & 0.965050 & 0.964505 & 0.0156171 & 0.0156171 \\
(I) & 0.967893 & 0.968325 & 0.0156501 & 0.0156516 \\
(II) & 0.963832 & 0.963599 & 0.0156030 & 0.0156035 \\
(III) & 0.966830 & 0.966288 & 0.0156377 & 0.0156378 \\
(IV) & 0.968110 & 0.968763 & 0.0156526 & 0.0156545 \\
\specialrule{0.1em}{1pt}{0pt}
\end{tabular}
\end{table}
\FloatBarrier
Across these points the maximum differences are
$|\Delta n_s|=6.53\times10^{-4}$ and
$|\Delta r|=1.85\times10^{-6}$.  The direct calculation therefore
supports the local-index approximation at the accuracy needed for the
parameter-space discussion here.  A dense numerical likelihood scan
would still be required for a precision parameter inference.

\section{Conclusions}
We have developed an analytical construction of exact single-field
inflationary backgrounds without imposing the slow-roll approximation.
The method maps the inflaton equation to an Abel equation of the first
kind and selects a functional form that renders the transformed system
exactly solvable.  The resulting class contains solutions with a
constant Hubble slow-roll parameter ($\eta_H$) and solutions with a
constant second Hubble-flow parameter ($\epsilon_2$).  The exact family
includes non-slow-roll evolution, although the regions selected by the
current joint CMB contour have small Hubble-flow parameters and lie
close to the slow-roll regime.

We derived a linear local criterion for the attraction of the exact
rolling backgrounds.  The same coefficient controls the local
evolution of the derivative of the time-dependent superhorizon mode,
\(
\mathcal{A}=\epsilon_1-\epsilon_2-3
\), so that $\mathcal{A}<0$ provides both restrictions used in our
scan.  The finite $\alpha=0$, $\lambda<0$ branch and all four displayed
$\alpha\neq0$ branches contain segments inside the adopted 95\% joint
CMB contour.  These results establish linear local attraction and
local superhorizon decay, but do not constitute a proof of global
nonlinear stability.

We also tested the local constancy of the scalar and tensor effective
indices.  Over all displayed 95\%-compatible regions, the logarithmic
rates $|d\ln Q_R/d\ln|\tau||$ and
$|d\ln Q_T/d\ln|\tau||$ are below $1.9\times10^{-3}$.  More
importantly, direct Mukhanov--Sasaki evolution at representative points
gives maximum shifts of only $6.53\times10^{-4}$ in $n_s$ and
$1.85\times10^{-6}$ in $r$ relative to the local-index estimates.
Normalizing the branch-(II) representative point to the measured scalar
amplitude fixes $V_0$ and gives an inflationary energy scale
$V_*^{1/4}\simeq1.14\times10^{16}\,\mathrm{GeV}$.

The e-fold analysis further separates the compatible solutions.  The
representative finite $\alpha=0$ point reaches $\epsilon_1=1$ only
after $N_*\simeq210$.  For $\alpha\neq0$, branches (I) and (IV) reach
$\epsilon_1=1$ with $N_*<50$, whereas branch (III) approaches
$\epsilon_1=-b<1$ and never crosses $\epsilon_1=1$.  By contrast,
branch (II) with $b=0.003$ has a nonempty intersection of the joint
$(n_s,r)$, local-attractor, superhorizon, and e-fold conditions.  The
representative point $\epsilon_{1*}=0.001$, $c_0=3$ gives
$N_*\simeq54.34$.  Thus, among the four displayed $\alpha\neq0$
branches, only branch (II) reaches $\epsilon_1=1$ within
$50\leq N_*<60$.  This statement concerns the end of accelerated
expansion only: no reheating sector is specified, so a transition to a
radiation-dominated era has not been demonstrated.

Our use of the public joint contour is more consistent than a mixture
of one-dimensional marginal cuts, but it is still not a full
model-specific likelihood analysis.  A dense likelihood-level scan,
and extensions to signatures such as primordial non-Gaussianity, are
left for future work.

\section*{Acknowledgments}
\begingroup
\setlength{\emergencystretch}{1em}
Jia-Wei Zhang was supported by NSFC under Contract No.~12275036, CQCSTC under Contract No. CSTB2025NSCQ-GPX0945 and No.~cstc2021jcyj-msxmX0681.
Yao Yu was supported in part by NSFC (Contracts No.~11905023, No.~12047564, and No.~12147102), the Natural Science Foundation of Chongqing (CQCSTC) under Contract No.~cstc2020jcyj-msxmX0555, and the Science and Technology Research Program of Chongqing Municipal Education Commission (STRPCMEC) under Contracts No.~KJQN202200605 and No.~KJQN202200621. Han Zhang and Bai-Cian Ke were supported in part by the National Natural Science Foundation of China (NSFC) under Contracts No.~11875054 and No.~12192263, the Joint Large-Scale Scientific Facility Fund of the NSFC and the Chinese Academy of Sciences under Contract No.~U2032104, and the Excellent Youth Foundation of the Henan Scientific Committee under Contract No.~242300421044.
Dong-Ze He was supported by NSFC under Contract No.~12275037 and STRPCMEC under Contract No.~KJQN202300609.
\par
\endgroup

\end{document}